\documentclass{llncs}
\usepackage{ifthen}
\usepackage{amsmath} 
\usepackage{amssymb} 
\usepackage{graphicx}
\usepackage{subcaption} 
\newcommand{\eg}{e.\,g.~}
\newcommand{\ie}{i.\,e.~}
\newcommand{\cf}{cf.~}
\DeclareMathOperator*{\argmin}{arg\,min}

\usepackage{tikz}
\usetikzlibrary{arrows, shapes, positioning, calc, automata}
\newcommand{\MarkRightAngle}[4][.6cm]
{\coordinate (tempa) at ($(#3)!#1!(#2)$);
	\coordinate (tempb) at ($(#3)!#1!(#4)$);
	\coordinate (tempc) at ($(tempa)!0.5!(tempb)$);
	\draw (tempa) -- ($(#3)!2!(tempc)$) -- (tempb);
}
\newcommand{\xa}{\vec{x}_a}
\newcommand{\ua}{\vec{u}_a}

\newcommand{\xb}{\vec{x}_b}
\newcommand{\ub}{\vec{u}_b}

\newcommand{\trans}{\vec{T}}

\newcommand{\transi}{\trans^{-1}}
\newcommand{\rfl}{\vec{F}}
\newcommand{\rflp}{\rfl^\prime}
\newcommand{\fourbyfour}{4\times4}
\newcommand{\pthree}{\mathbb{P}^3}
\newcommand{\rfourfour}{\mathbb{R}^{\fourbyfour}}

\newcommand{\radon}{\mathcal{R}} 

\newcommand{\tft}{\trans\rfl\transi}

\begin{document}

\title{Double Your Views -- Exploiting Symmetry in Transmission Imaging}
\titlerunning{Double Your Views}

\author{Alexander~Preuhs\inst{1} \and Andreas~Maier~\inst{1}\and Michael~Manhart\inst{2} \and Javad~Fotouhi\inst{3} \and Nassir~Navab\inst{3} \and Mathias~Unberath\inst{3}}
\authorrunning{Preuhs et al.} 
\institute{Pattern Recognition Lab, Friedrich-Alexander-Universit{\"a}t Erlangen-N{\"u}rnberg\\
	\and
Siemens Healthcare GmbH, Forchheim, Germany\\
	\and
Computer Aided Medical Procedures, Johns Hopkins University}

\maketitle              

\begin{abstract}
For a plane symmetric object we can find two views~---~mirrored at the plane of symmetry~---~that will yield the exact same image of that object. In consequence, having one image of a plane symmetric object and a calibrated camera, we can automatically have a second, virtual image of that object if the \mbox{3-D} location of the symmetry plane is known.
In this work, we show for the first time that the above concept naturally extends to transmission imaging and present an algorithm to estimate the \mbox{3-D} symmetry plane from a set of projection domain images based on Grangeat's theorem. We then exploit symmetry to generate a virtual trajectory by mirroring views at the plane of symmetry. If the plane is not perpendicular to the acquired trajectory plane, the virtual and real trajectory will be oblique. The resulting X-shaped trajectory will be data-complete, allowing for the compensation of in-plane motion using epipolar consistency. We evaluate the proposed method on a synthetic symmetric phantom and, in a proof-of-concept study, apply it to a real scan of an anthropomorphic human head phantom.
\end{abstract}
\section{Introduction}
Symmetry is a powerful concept with applications ranging from art to physics and mathematics~\cite{field2009symmetry}. This manuscript is concerned with symmetry in computer vision where we consider a theoretically sound yet surprisingly little known property of symmetric objects: when imaging a symmetric object using a calibrated camera, knowledge of the \mbox{3-D} symmetry plane yields a second, virtual camera that corresponds to a mirrored version of the image seen by the true camera. This circumstance enables metric \mbox{3-D} stereo reconstruction of symmetric objects using a single calibrated camera~\cite{rothwell1993extracting,franccois2002reconstructing,franccois2003mirror}.

For the first time, we demonstrate that the above property naturally extends to transmission imaging, \ie X-ray fluoroscopy, and devise image-based algorithms that exploit this circumstance to estimate intra-scan motion in circular C-arm cone-beam computed tomography (CBCT). In CBCT imaging, all camera positions are calibrated suggesting that a virtual source trajectory becomes available once the \mbox{3-D} symmetry plane is known. We show that 1) this plane can be estimated efficiently from multiple projective images and 2) circular trajectories in a plane oblique to the plane of symmetry contain information that substantially benefits motion detection using recent consistency conditions.
\section{Methods}
\subsection{Epipolar Consistency Conditions}
\paragraph{Theory:}
\begin{figure}[b]
	\begin{center}
		\begin{tikzpicture}[xscale=0.65,yscale=0.55]
		\input{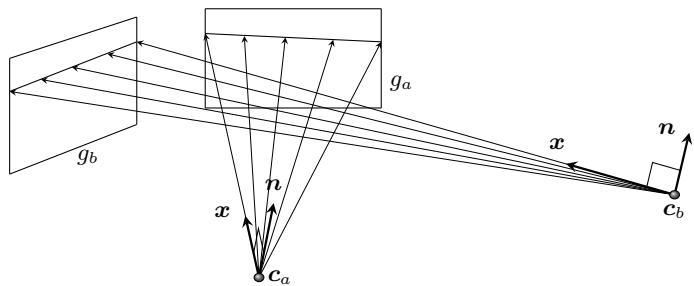}
		\end{tikzpicture}
		\caption{Schematic drawing of a scene including two projections.}
		\label{fig:epipolarGeometry}
	\end{center}
\end{figure}
In CBCT an X-ray source radially emits photons, that~---~after attenuation~---~are registered at a detector. The attenuation process for a ray is described by an integral. However, due to the radial structure of the rays, integrating along a detector line does not result in the Cartesian plane integral of the underlying object $f$ but differs by a radial weighting. 

Grangeat's theorem describes the connection between this weighted integral and a plane integral~---~\ie the \mbox{3-D} radon value $\radon f(\vec{n},d)$ describing the integral along a plane with normal $\vec{n} \in \mathcal{S}^2$ at distance $d$. Using a derivative operation the radial weighting can be canceled out.
Grangeat defined an intermediate function $S_\lambda(\vec{n})$ that is calculated from  projection data and is related to the derivative of the \mbox{3-D} radon transform
\begin{equation}
	S_\lambda(\vec{n})=
	\int_{\mathcal{S}^2} \delta'(\vec{x}^\top \vec{n})  g_\lambda(\vec{x})  d \vec{x}=
	\frac{\partial}{\partial d}  \radon f(\vec{n},d)\lvert_{d=\vec{c}^\top_\lambda\vec{n}}
	\enspace,
	\label{eq:intermediate}
\end{equation}
where $\delta'(\cdot)$ describes the derivative of the Dirac delta distribution,
 $g_\lambda(\vec{x})$ describes a single value on the detector with $\lambda$ the projection index, $\vec{c}_\lambda$ the source position and $\vec{x}$ a vector from the source to a detector pixel. The geometry for two projections is visualized in Fig.\,\ref{fig:epipolarGeometry}. A detailed derivation of Eq.~\eqref{eq:intermediate} can be found in~\cite{Defrise1994}, and some simplifications are discussed in~\cite{Aichert2015}. 
From Eq.~\eqref{eq:intermediate} it directly follows that two projections $a, b$ must satisfy
\begin{align}
S_a(\vec{n}) =
S_b(\vec{n})
&& \forall \vec{n} \in \mathcal{S}^2: \vec{c}_b^\top \vec{n} = \vec{c}_a^\top \vec{n} 
\enspace.
\label{eq:consistency}
\end{align}
If the geometric calibration is wrong, \eg due to object motion, Eq.~\eqref{eq:consistency} will not hold. Thus, we can use it as a measure of inconsistency. 
The global indexing by the plane normal $\vec{n}$ can be replaced by a local projection-pair-dependent indexing per the epipolar geometry. This particular sampling  of the intermediate function is commonly denoted as epipolar consistency (EC) \cite{Aichert2015} and allows to efficiently evaluate redundant values only based on the corresponding projection matrices $\vec{P}_a$ and $\vec{P}_b$, allowing the reformulation of Eq. \eqref{eq:consistency} to
\begin{align}
\vec{S}_a(\vec{P}_a, \vec{P}_b) = \vec{S}_b(\vec{P}_b, \vec{P}_a) 
\enspace,
\label{eq:epipolarConsistency}
\end{align} 
where $\vec{S}$ denotes the array of intermediate values computed from $S$.
\paragraph{Short Scans and the Circular Trajectory:}
State of the art head imaging protocols for CBCT consist of a circular trajectory of 496 projections. Hereby the source detector gantry rotates around the patient covering a $200^\circ$ segment. All source positions within the trajectory lie on a plane, typically referred to as trajectory plane. We define this plane coincident with the $\vec{x},\vec{y}$-plane.

Rigid patient motion or geometry misalignment can be estimated and compensated for using EC \cite{Frysch2015,Maass2014,Preuhs2018}. This is achieved by finding a rigid transformation $\trans_i$ for each projection matrix $\vec{P}_i$, accounting for the patient motion at acquisition of projection $i$. The motion is expected to be compensated, when the inconsistency between all projections is minimal. The result of motion estimation is a set of rigid transformations $\trans = [\trans_1,\dots\trans_{496}]$ that satisfy
\begin{equation}
\hat{\vec{T}} = \argmin_{\vec{T}} \sum_{a,b = 1}^{496} \left\lVert\vec{S}_a(\vec{P}_a\vec{T}_a,\vec{P}_b\vec{T}_b) - \vec{S}_b(\vec{P}_b\vec{T}_b,\vec{P}_a\vec{T}_a)\right\lVert_2
\enspace.
\label{eq:ECCOptimization}
\end{equation}

However, there are theoretical limitations due to the geometry of the circular trajectory. Most radon planes of the object that include two source positions are almost parallel to the trajectory plane. As the consistency is based on the radon value, only motion that steps out of the radon plane is detectable. As a consequence only out-plane motion ($r_x, r_y, t_z$) can be estimated well, while in-plane motion ($t_x, t_y, r_z$) cannot be estimated robustly \cite{Frysch2015}. In the following, we show that we can exploit symmetry to generate a short scan-like trajectory which is data complete and, thus, beneficial for estimation of in-plane motion.

\subsection{Symmetric View Augmentation}

\paragraph{Symmetry in Transmission Imaging:}
\begin{figure}
	\begin{center}
		\begin{tikzpicture}[xscale=0.8,yscale=0.6]
		\input{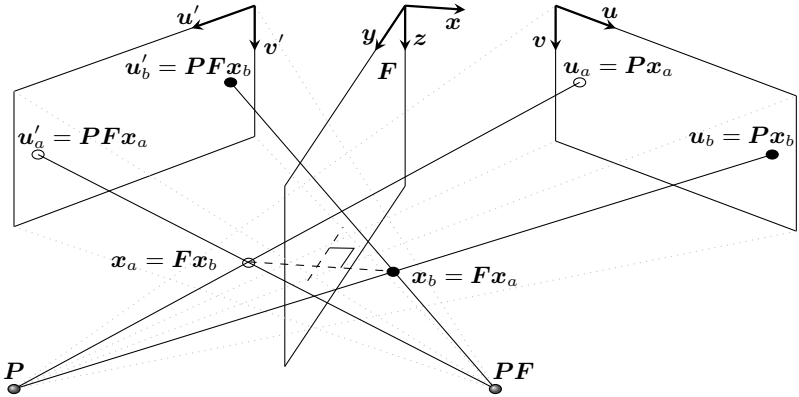}
		\end{tikzpicture}
		\caption{Visualization of a plane symmetric scene.}
		\label{fig:symmetry}
	\end{center}
\end{figure}
Consider a plane-symmetric scene as depicted in Fig.\,\ref{fig:symmetry}. The two points $\xa \in \pthree$ and $\xb \in \pthree$ are symmetric to the $\vec{z},\vec{y}$-plane. This bilateral symmetry relation can be expressed by an involutive isometric transformation $\rfl$~---~\ie a reflection matrix~---~as
\begin{align}
\xa = \rfl \, \xb && \xb = \rfl \, \xa \enspace ,
\label{eq:mirror_point}
\end{align}
where $\rfl \in \rfourfour$ only flips the sign of the $x$ component. 
Since $\rfl$ is an isometric transformation we find the mirrored projection matrix as $\vec{P}\rfl$ (\cf Fig.\,\ref{fig:symmetry}). The resulting image on the right detector  will be the projection of the points $\xa$ and $\xb$ under $\vec{P}$ and the resulting image on the left detector will be the projected points $\xa$ and $\xb$ under $\vec{P}\rfl$ giving
\begin{align}
\vec{u}_a = \vec{P} \xa && \vec{u}_b = \vec{P} \xb && \vec{u}_a^\prime = \vec{P}\rfl \xa && \vec{u}_b^\prime = \vec{P}\rfl \xb \enspace.
\label{eq:proj_points}
 \end{align}
Inserting the symmetry relation given by Eq.~\eqref{eq:mirror_point} in the two leftmost equations of Eq.~\ref{eq:proj_points} gives the relation 
\begin{align}
\vec{u}_a = \vec{P} \xa = \vec{P}\rfl \xb= \vec{u}_b^\prime  && \vec{u}_b = \vec{P} \xb  = \vec{P}\rfl \xa= \vec{u}_a^\prime \enspace.
\label{eq:proj_points2}
\end{align}
This result allows to conclude that both detector images will exhibit the exact same image. Note that this only holds since the reflection of the projection matrix also flips the $\vec{u}$ and $\vec{v}$ axis. Consequently, a transmission image of a plane symmetric object can be interpreted as acquired under either the projection $\vec{P}$ or $\vec{P}\rfl$. This observation allows to effectively double the views of an acquisition if the symmetry plane is known.
\paragraph{Symmetry Plane Estimation:} 
We apply EC (\cf Eq.~\eqref{eq:ECCOptimization}) to find the reflection $\rflp$ that represents the most consistent transformation, which is~---~by definition~---~the reflection at the symmetry plane. To optimize for a certain symmetry plane, we need to find the transformation describing $\rflp$, which is given by $\rflp = \tft$, with $\trans$ being a rigid transformation. The reflection $\rflp$ is then found by optimizing for $\hat{\vec{\trans}}$ minimizing the inconsistency defined as
\begin{equation}
\hat{\vec{\trans}} = \argmin_{\vec{\trans}} \sum_{a,b = 1}^N \left\lVert \vec{S}_a(\vec{P}_a\tft,\vec{P}_b) - \vec{S}_b(\vec{P}_b,\vec{P}_a\tft) \right\rVert_2
\enspace,
\label{eq:optimizeSymmetryPlane}
\end{equation} 
where $N$ is the number of projections used for finding the symmetry plane.
\paragraph{The X-Trajectory:}
\begin{figure}
	\begin{center}
		\begin{tikzpicture}[xscale=0.8,yscale=0.6]
		\input{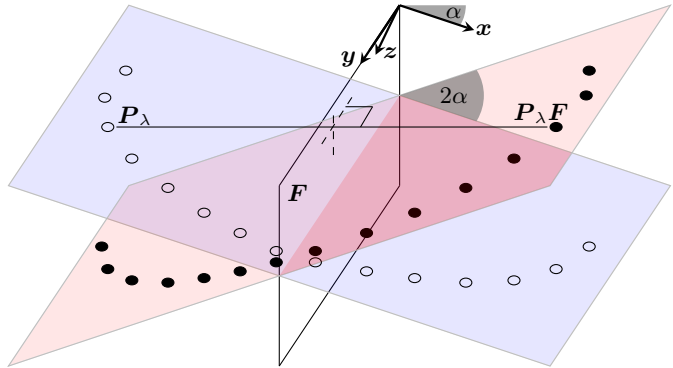}
		\end{tikzpicture}
		\caption{Visualization of X-trajectory. The acquired trajectory is embedded in the blue trajectory plane (blank dots) and the mirrored virtual trajectory is embedded in the red trajectory plane (solid dots).}
		\label{fig:Xtrajectory}
	\end{center}
\end{figure}
If the symmetry plane of the scanned object is oblique to the trajectory plane of a short scan by an angle $\alpha$ as visualized in Fig.\,\ref{fig:Xtrajectory}, the mirrored trajectory plane will be rotated to the acquired trajectory plane by $2\alpha$. Thus, for adequate angles $\alpha$, the combined trajectory fulfills Tuy's condition and the short scan becomes data complete. This in turn enables the use of Grangeat's theorem to detect in-plane motion.
\subsection{Experiments}
\paragraph{Data:}
\begin{figure}[b]
	\centering
	\begin{subfigure}{0.24\textwidth}
		\raisebox{-\height}{\includegraphics[width=\textwidth]{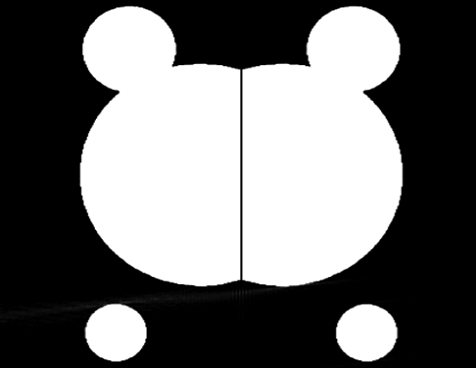}}
		\caption{}
		\label{subfig:phantomVolume}
	\end{subfigure}
	\begin{subfigure}{0.24\textwidth}
		\raisebox{-\height}{\includegraphics[width=0.48\textwidth]{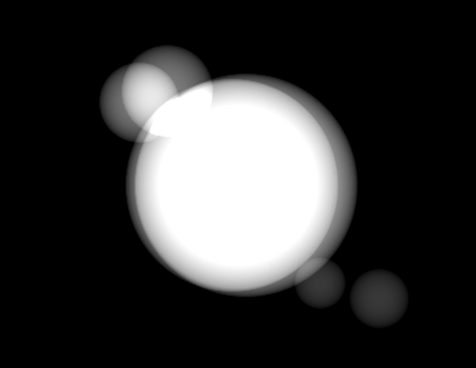}}
		\raisebox{-\height}{\includegraphics[width=0.48\textwidth]{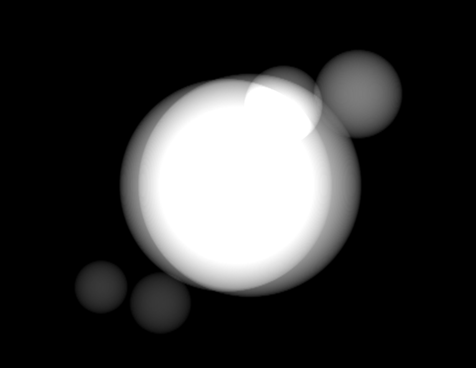}}%
		\vspace{.4ex}
		\raisebox{-\height}{\includegraphics[width=0.48\textwidth]{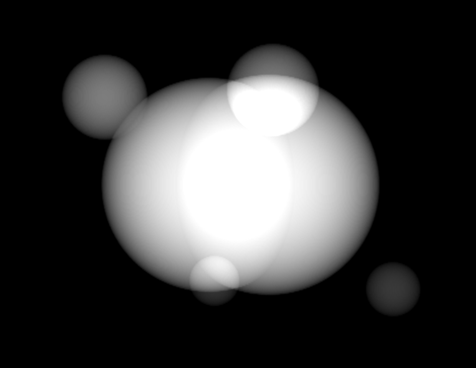}}
		\raisebox{-\height}{\includegraphics[width=0.48\textwidth]{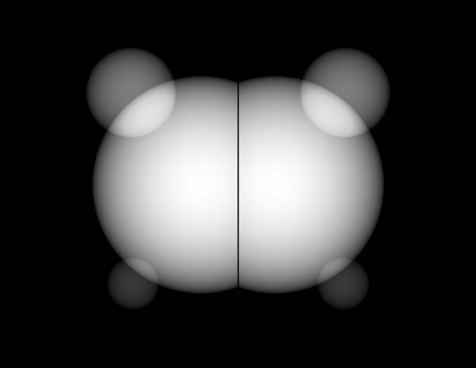}}
		\caption{}
		\label{subfig:phantomDRR}
	\end{subfigure}	
	\hspace{0.4ex}
	\begin{subfigure}{0.24\textwidth}
		\raisebox{-\height}{\includegraphics[width=\textwidth]{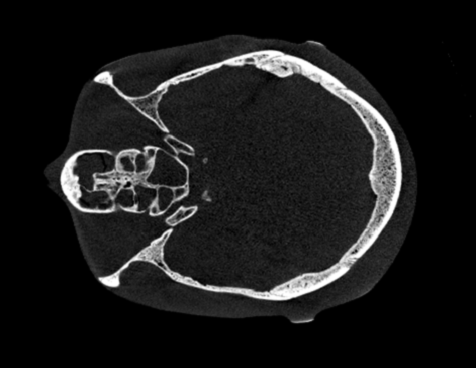}}
		\caption{}
		\label{subfig:headVolume}
	\end{subfigure}
	\begin{subfigure}{0.24\textwidth}
		\raisebox{-\height}{\includegraphics[width=0.48\textwidth]{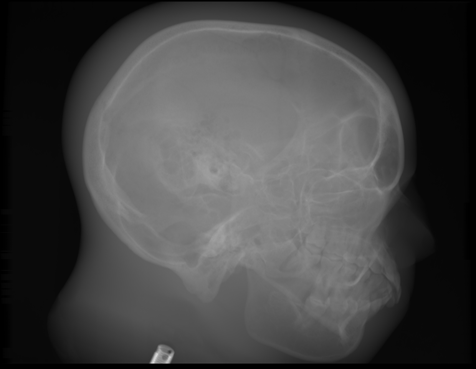}}
		\raisebox{-\height}{\includegraphics[width=0.48\textwidth]{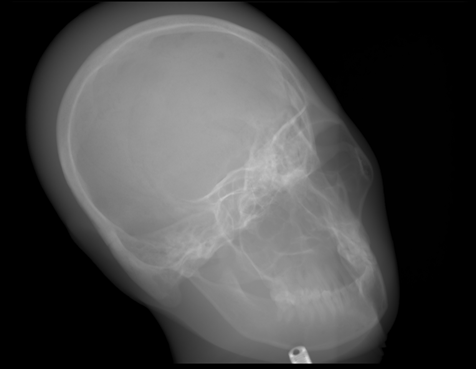}}%
		\vspace{.4ex}
		\raisebox{-\height}{\includegraphics[width=0.48\textwidth]{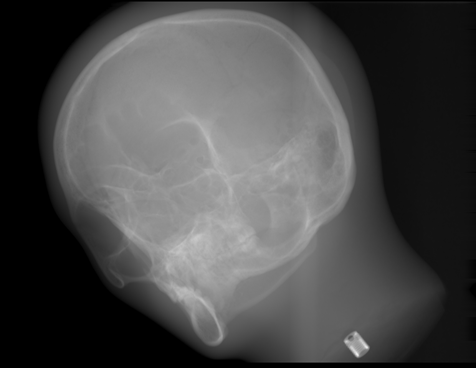}}
		\raisebox{-\height}{\includegraphics[width=0.48\textwidth]{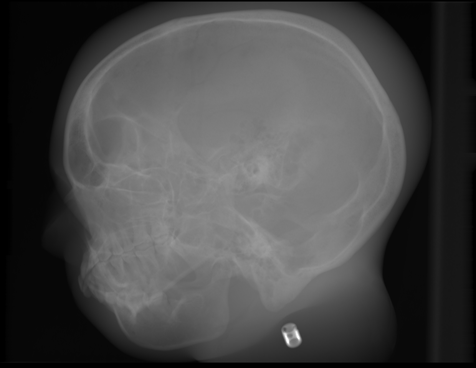}}
		\caption{}
		\label{subfig:headScan}
	\end{subfigure}
	\caption{(a): Slice through the symmetric phantom. (b): Digitally rendered radiographs (DRR) of the phantom. (c): Aligned reconstruction of anthropomorphic head phantom. (d): raw projection data from a short scan of the head phantom.}
\end{figure}
To evaluate our method we synthetically generated a plane symmetric phantom, consisting of four small balls, and two half spheres (\cf Fig. \ref{subfig:phantomVolume}). From this phantom, a short scan is simulated \emph{in silico} (\cf Fig.\,\ref{subfig:phantomDRR}).
The second dataset is a short scan acquired from a real anthropomorphic human head phantom using a robotic C-arm system (Artis zeego, Siemens Healthcare GmbH, Germany). The phantom is placed, such that the expected symmetry plane is oblique to the trajectory plane. As expected in a real clinical case, the head phantom does not exhibit a perfect plane symmetry. A slice through the reconstruction and projections from the acquired short scan of the head are shown in Figs.~\ref{subfig:headVolume} and \ref{subfig:headScan}, respectively.

\paragraph{Estimation of Symmetry Plane:}
Using the synthetic and real datasets, we first estimate the symmetry plane. The plane is found from projection domain images only by minimizing Eq. \eqref{eq:optimizeSymmetryPlane} using the Nelder-Mead method. The optimization searches for the symmetry plane parameters described by three DoF.

\paragraph{Application to Rigid Motion:}
To study the impact of the X-trajectory in dependence of $\alpha$ on in- and out-plane motion, we add a rigid spline motion to each motion parameter. The motion amplitude is in the range of $\pm$ $0.3$ mm or degree, respectively, and distributed only in the central part of the trajectory, where no opposing views are available. Then we compute combined consistency grids of the motion affected trajectory using the synthetic phantom. The grid is build up by a ($N${}$\times${}$N$) matrix $\vec{C}$, where each element $c_{ab}$ denotes the consistency between views $a$ and $b$. The lower-left triangle of the consistency grid denotes the conventional EC (CEC) computed from two views on the circular trajectory. The upper-right triangle is computed as the EC between an acquired view $a$ and a mirrored view $b$, which we denote as mirrored EC (MEC).

In addition we inspect the inconsistency induced by an $r_z$ motion pulse distributed over view $238-288$  on the acquired human head phantom consisting of $496$ views. We compare the CEC solely computed from the short scan, epipolar consistency between mirrored and acquired view (MEC) and a combination of both (combined CEC and MEC).   
\section{Results and Discussion}
\paragraph{Estimation of Symmetry Plane:}
\begin{figure}[b]
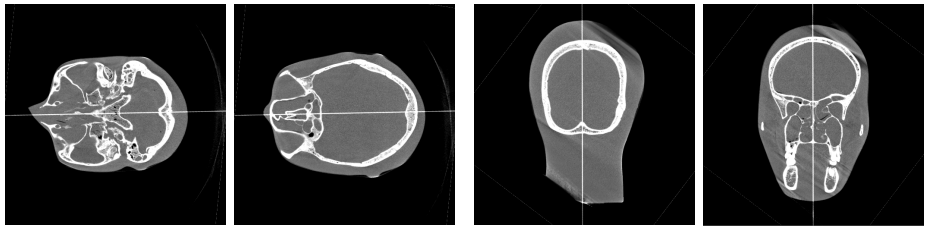

	\include{fig_sym_plane_estimation}
	\caption{Reconstruction of acquired head phantom. The volume is aligned to the symmetry plane (white line) and shown from an axial and coronal view.}
	\label{fig:reco_sym}
\end{figure}
The estimated and ground truth symmetry plane parameters for the synthetic phantom are listed in Tab.~\ref{tab:symplane}. Estimation succeeded with very high accuracy.
\begin{table}[h]
	\centering
    \caption{Ground truth symmetry plane parameters (normal and signed distance from origin) of synthetic phantom and estimated symmetry plane parameters.}
\begin{tabular*}{0.93\textwidth}{p{2.5cm} p{2.3cm} p{2.3cm} p{2.3cm} p{2.3cm}}
	\hline 
	& $n_x$ & $n_y$ & $n_z$ & $d$ \\ 
	\hline 
	Ground Truth & 0 & 1 & 0 & 0 \\ 
	Estimation & 0.0000926 & 0.9999999 & 0.0000005 & 0.0000822 \\ 
	\hline 
\end{tabular*} 
\label{tab:symplane}
\end{table}
The head phantom, while not perfectly symmetric, exhibits a well defined symmetry plane that was estimated very robustly. A reconstruction aligned w.r.t. the symmetry plane is depicted in Fig.\,\ref{fig:reco_sym}.

\begin{figure}[tb]
	\input{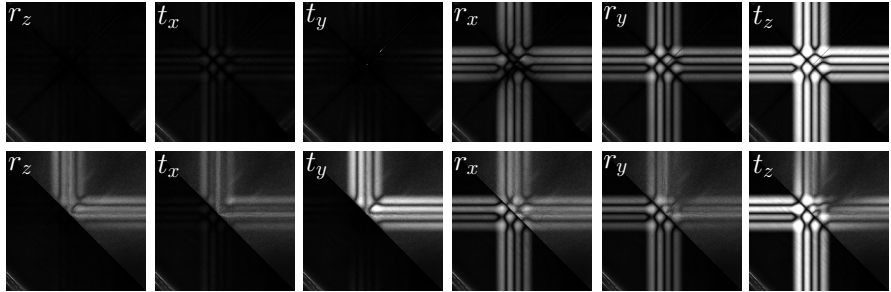}
	\caption{Inconsistency due to motion in the trajectory using the synthetic phantom. Bright pixels encode a high inconsistency and dark regions encode consistent view pairs. Upper row:  $\alpha = 0^\circ$, lower row: $\alpha =30^\circ$.}
	\label{fig:Xangle_synth}
\end{figure}

\paragraph{Application to Rigid Motion:} 
Comparing upper and lower row of Fig.\,\ref{fig:Xangle_synth} that correspond to $\alpha = 0^\circ$ and $\alpha =30^\circ$, respectively, the impact of the X-trajectory is evident. The CEC (lower left triangle of the grid) detects inconsistency within out-plane parameters (three rightmost columns) while in-plane motion (three leftmost columns) is not detected well. Using MEC and an angle $2\alpha=60^\circ$ between the acquired and mirrored trajectory plane, prominently reveals in-plane motion.  
\begin{figure}[t]
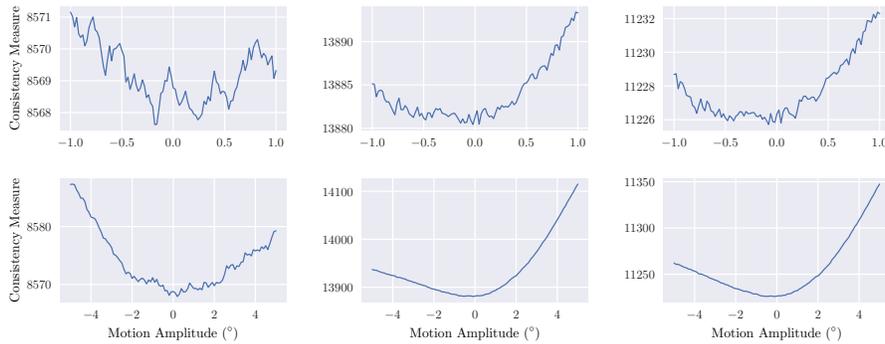

		\begin{subfigure}{0.32\textwidth}
		\resizebox{\linewidth}{!}{
			\input{tikz/convent1}}
		\end{subfigure}
		\begin{subfigure}{0.32\textwidth}
			\resizebox{\linewidth}{!}{
			\input{tikz/symmetry1}}
		\end{subfigure}	
		\begin{subfigure}{0.32\textwidth}
			\resizebox{\linewidth}{!}{
			\input{tikz/combined1}}
		\end{subfigure}
	
	\begin{subfigure}{0.32\textwidth}
		\resizebox{\linewidth}{!}{
			\input{tikz/convent5}}
	\end{subfigure}
	\begin{subfigure}{0.32\textwidth}
		\resizebox{\linewidth}{!}{
			\input{tikz/symmetry5}}
	\end{subfigure}	
	\begin{subfigure}{0.32\textwidth}
		\resizebox{\linewidth}{!}{
			\input{tikz/combined5}}
	\end{subfigure}
\caption{Sensitivity of consistency measure to motion impulse. Left column: CEC. Middle column: MEC. Right column: combined CEC and MEC.}
\label{fig:rzcomp}
\end{figure}

Figure\,\ref{fig:rzcomp} shows the different consistency measures (CEC, MEC, combined CEC and MEC) responding to a motion impulse on the acquired data. All measures are able to detect large scale motion. However, using CEC the global optimum is displaced by $0.2^\circ$. The MEC optimum is displaced by $0.03^\circ$, combined CEC and MEC shows a displacement of $0.06^\circ$, suggesting that, in this setup, MEC is more accurate than CEC by a factor of 10.

\section{Conclusion}
We presented the concept of plane symmetry for transmission imaging and provided an algorithm to estimate the \mbox{3-D} plane of symmetry based on projection images only. In combination with a short scan trajectory oblique to the symmetry plane, an X-shaped trajectory arises that is associated with several benefits. For adequate angles $\alpha$, both in- and out-of-plane motion directions are detectable using Grangeat's theorem. This property naturally arises from the observation that the X-trajectory is Tuy-complete. We have evaluated the proposed algorithm on a real scan of an anthropomorphic head phantom. Despite being only partially symmetric, the proposed concept of exploiting symmetry was still found applicable. Future research is needed to find effective optimization strategies to estimate complex motion patterns. We conclude that symmetry is a powerful concept in transmission imaging with the potential to benefit diverse imaging problems that make use of consistency condition such as calibration, beam hardening- and truncation-correction.
\paragraph{}
~\\
\textbf{Disclaimer:} The concepts and information presented in this paper are based on research and are not commercially available.
\bibliographystyle{splncs}
\bibliography{main}

\end{document}